# Evidence of Ising pairing in superconducting NbSe$_2$ atomic layers

Xiaoxiang Xi[1]*, Zefang Wang[1]*, Weiwei Zhao[1], Ju-Hyun Park[2], Kam Tuen Law[3], Helmuth Berger[4], László Forró[4], Jie Shan[1δ], and Kin Fai Mak[1δ]

[1]Department of Physics, The Pennsylvania State University, University Park, Pennsylvania 16802-6300, USA

[2]National High Magnetic Field Laboratory, Florida State University, Tallahassee, Florida 32310, USA

[3]Department of Physics, Hong Kong University of Science and Technology, Clear Water Bay, Hong Kong, China

[4]Institute of Condensed Matter Physics, Ecole Polytechnique Fédérale de Lausanne, 1015 Lausanne, Switzerland

*These authors contributed equally to this work.

δCorrespondence to: jus59@psu.edu, kzm11@psu.edu

**Two-dimensional transition metal dichalcogenides with strong spin-orbit interactions and valley-dependent Berry curvature effects have attracted tremendous recent interests[1-7]. Although novel single-particle and excitonic phenomena related to spin-valley coupling have been extensively studied[1,3-6], effects of spin-momentum locking on collective quantum phenomena remain unexplored. Here we report an observation of superconducting monolayer NbSe$_2$ with an in-plane upper critical field over six times of the Pauli paramagnetic limit by magneto-transport measurements. The effect can be understood in terms of the competing Zeeman effect and large intrinsic spin-orbit interactions in non-centrosymmetric NbSe$_2$ monolayers, where the electronic spin is locked to the out-of-plane direction. Our results provide a strong evidence of unconventional Ising pairing protected by spin-momentum locking and open up a new avenue for studies of non-centrosymmetric superconductivity with unique spin and valley degrees of freedom in the exact two-dimensional limit.**

Monolayer transition metal dichalcogenide (TMD) of the hexagonal structure consists of a layer of transition metal atoms sandwiched between two layers of chalcogen atoms in the trigonal prismatic structure[8] (Fig. 1a). It possesses out-of-plane mirror symmetry and in-plane inversion asymmetry. The presence of the transition metal also gives rise to large spin-orbit interactions (SOIs). The mirror symmetry restricts the crystal field ($\vec{\varepsilon}$) to the plane. The SOIs split the spin states at finite momentum $\vec{k}$ in the absence of inversion symmetry. They manifest as an effective magnetic field along the direction of $\vec{k} \times \vec{\varepsilon}$, which is out-of-plane for the restricted two-dimensional (2D) motion of electrons in the plane. The electronic spin is thus oriented in the out-of-plane direction and in opposite directions for electrons of opposite momenta[1-3] (Fig. 1a). Such spin-momentum locking is destroyed in the bulk where inversion symmetry and spin degeneracy are restored[1,2,7] (Fig. 1b). Novel valley- and spin-dependent phenomena including optical orientation of the valley polarization[3,4] and the valley Hall effect[5] arisen from spin-momentum locking have been recently demonstrated in group-VI TMD monolayers such as MoS$_2$. The effect of spin-momentum locking on collective quantum phenomena, however, remains experimentally unexplored.

In this work, we demonstrate very high in-plane upper critical fields induced by spin-momentum locking in *superconducting monolayers* of group-V TMD niobium diselenide (NbSe$_2$). This was achieved through transport and magneto-transport studies of high quality NbSe$_2$ devices of varying layer thickness. Bulk 2H-NbSe$_2$ is a well-studied type II anisotropic multiband superconductor with a zero-field critical temperature $T_{C0} \approx$ 7 K[9-14]. Superconductivity in atomically thin NbSe$_2$[15,16] down to the monolayer thickness[17] has also been observed recently. Monolayer NbSe$_2$ can be viewed as heavily hole doped monolayer MoSe$_2$[8,18]: The Fermi surface is composed of one pocket at the Γ point and two pockets at the K and the K' points of the Brillouin zone with each pocket spin split into two pockets by the SOIs[18] (Fig. 1a); The spin splitting at the Γ point is much smaller than that at the K (K') point[18] and the spin-momentum locking effects are dominated by the K (K') pockets.

In conventional superconductors, superconductivity can be quenched under a sufficiently high external magnetic field by the orbital[11,19] and spin effect[20,21], which are originated, respectively, from the coupling between the electron momentum and the magnetic field (e.g. vortex formation) and from spin alignment by the magnetic field. In the limit of monolayer thickness, the interlayer coupling vanishes and the orbital effect is absent for an in-plane magnetic field $H_\parallel$. The in-plane upper critical field (critical field at zero temperature) $H_{c20}^\parallel$ is determined by direct spin-flip, known as the Pauli paramagnetic limit $H_P \approx 1.84\, T_{C0}$ (in Tesla for $T_{C0}$ in Kelvin) for isotropic BCS superconductors[20,21]. Under this field the Zeeman splitting energy matches the superconducting energy gap, or equivalently, the binding energy of a Cooper pair. Spin-momentum locking and the consequent pairing of the K and K' electrons with spin locked to the two opposite out-of-plane directions (Fig. 1a), referred to as *Ising pairing*, is expected to enhance $H_{c20}^\parallel$ significantly above the Pauli limit $H_P$ in monolayer NbSe$_2$.

In our experiment, we fabricated monolayer and few-layer samples of NbSe$_2$ by mechanical exfoliation of 2H-NbSe$_2$ single crystals followed by direct transfer onto SiO$_2$/Si substrates with pre-patterned electrodes and encapsulation by thin layers of hexagonal boron nitride (see Methods). The sample thickness was determined with monolayer accuracy by optical spectroscopy, particularly, the frequency shift of the interlayer shear mode[17]. The crystal symmetry is characterized by optical second harmonic generation. (See Supplementary Materials Section 1 for details on optical characterization). Figure 1c is an optical image of a representative device (bilayer in this case). The temperature dependence of the normalized four-point resistance $R(T)/R(300\text{ K})$ at zero magnetic field is shown in figure 1d from 2 – 300 K for a typical bulk, bilayer and monolayer device. (See Supplementary Materials Section 2 for original data and more devices). All samples show a metallic behavior with phonon-limited transport at high temperature (with $R \propto T$) and disorder-limited transport at low temperature (with $R$ approaching a constant value) before reaching the superconducting state. The residual resistance ratio (RRR) evaluated using the room temperature resistance and the normal state resistance right above the superconducting transition $R_n$ (taken at 8 K) varies from ~

30 in the bulk to ~ 10 in the monolayer device. The square resistance per layer (at 8 K) was calculated to be ≈ 200 Ω for the bilayer (which has a good geometry) and similar values were estimated for other devices. These values are much smaller than $\frac{h}{4e^2} \approx 6450$ Ω, where a disorder-induced superconductor-insulator transition emerges[22]. Here $e$ is the electron charge and $h$ is the Planck constant. Our samples are therefore in the low disorder regime compared to similar layered compounds studied previously[16] and the disorder effects on $T_{C0}$ are expected to be small[22].

Figure 2a shows the temperature dependence near the transition of the resistance normalized by $R_n$ for NbSe$_2$ samples of varying thickness $N$. Superconductivity is observed for all samples down to the monolayer thickness. A significant drop in the transition temperature accompanied with a significant broadening is observed for $N < 4$ (~ 2.68 nm). The broadening can be attributed to enhanced thermal fluctuations in 2D[19,23] when the sample thickness falls below the bulk out-of-plane coherence length of 2.7 nm[10]. We have used the Aslamozov-Larkin formula[24] to determine $T_{C0}$ (solid lines, Fig. 2a), which is close to the temperature corresponding to $0.5R_n$. We have also performed current excitation measurements to investigate the importance of phase fluctuations, and the Berenzinskii-Kosterlitz-Thouless transition temperature[23] is found to be close to $T(0.01R_n)$. Figure 2b summarizes the $N$-dependence of $T_{C0}$, $T(0.5R_n)$ and $T(0.01R_n)$. The monotonic dependence of $T_{C0}$ on $N$ can be accounted for by the decreasing interlayer Cooper pairing[25,26] (the linear dependence of $[ln\frac{T_{C0}(N)}{T_{C0}(N=1)}]^{-1}$ on $[\cos(\frac{\pi}{N+1})]^{-1}$ in the inset of Fig. 2b). Effects on $T_{C0}$ from the competing charge-density-wave order that is enhanced with decreasing $N$ are thus likely to be weak in NbSe$_2$[14]. (See Supplementary Materials Section 3 and 4 for details on the study of the characteristics of 2D superconductivity in NbSe$_2$).

With the above understanding of the nature of superconductivity in 2D NbSe$_2$, we now study its magnetic response. Figure 3a-c show the temperature dependence of the four-point resistance for a bulk, trilayer and monolayer sample under both out-of-plane ($H_\perp$) and in-plane magnetic fields ($H_\parallel$). A second monolayer device with two-point measurements up to 20 T is also included. For all samples, we normalize the temperature by the corresponding $T_{C0}$, and the resistance by $R_n$. We define the critical temperature $T_C$ under a finite magnetic field $H_{c2}$ as the temperature corresponding to 50% of $R_n$, in accordance with the zero field convention. For two-point measurements, as shown in Fig. 3c, we assign $T_{C0}$ as the temperature at which a rapid resistance drop occurs while the sample is cooled from the normal state. It is clear that for all samples superconductivity is more susceptible to $H_\perp$ than to $H_\parallel$ and the magnetic anisotropy is significantly larger in atomically thin samples than in the bulk. (For angular dependence study of the magnetic response refer to Supplementary Materials Section 5).

We summarize the $H_{c2} - T_C$ phase diagram for differing sample thickness $N$ in figure 4 for both $H_\perp$ (open symbols) and $H_\parallel$ (filled symbols). For comparison, we normalize the critical field $H_{c2}$ by the BCS Pauli paramagnetic limit $H_P$ and the critical temperature $T_C$ by $T_{C0}$ for each sample. For out-of-plane fields, our experiment shows a linear $H_{c2}^\perp - T_C$ dependence that is largely thickness independent. The result can be explained by considering the orbital effect, i.e. overlaps of vortex cores, as the major quenching mechanism. $H_{c20}^{\perp}$ is determined by the combined effect of in-plane coherence length and transport mean free path[19]. While the coherence length increases with reducing thickness (because of the reduction in $T_{C0}$), the mean free path decreases as the material becomes more disordered. The net effect is a weak $H_{c2}^{\perp} - T_C$ dependence on $N$. The measured upper critical field $H_{c20}^{\perp} \approx 4$ T ($<< H_P$) agrees well with the value reported for bulk NbSe$_2$[9,10].

In contrast, for in-plane field, the $H_{c2}^{\parallel} - T_C$ dependence near $T_{C0}$ in the monolayer is much steeper than in the bulk and follows a square-root instead of a linear dependence. More significantly, the monolayer upper critical field $H_{c20}^{\parallel}$ far exceeds its Pauli limit $H_P$ while $H_{c20}^{\parallel} \gtrsim H_P$ in the bulk. We note that large enhancements of $H_{c20}^{\parallel}$ ($>>H_P$) have been observed in other systems[9,27-31]. Many mechanisms have been discussed including strong coupling[37,38], modified electron g-factor[19,30], interaction effects[35-37], remnant magnetic susceptibility at low temperature[30], Rashba SOI[30], and spin-orbit scattering from impurities[27,28,31-34]. These effects, however, are insignificant in 2D NbSe$_2$[11,12,37,39]. (See Supplementary Materials Section 6 for detailed discussions). In non-centrosymmetric monolayer NbSe$_2$, the large enhancement of $H_{c20}^{\parallel}$, as discussed above, can arise from the strong intrinsic SOIs and spin-momentum locking. In the absence of the orbital effect, $H_{c20}^{\parallel}$ is determined by the alignment of spin by the external field. The upper critical field can be estimated by noting that the in-plane component of the spin magnetic moment is reduced to $\sim \frac{H_\parallel}{H_{SO}} \mu_B$ due to the competing SOI and Zeeman effect. Here $H_{SO} \equiv \frac{\Delta_{SO}}{\mu_B}$ is the effective magnetic field from the SOI with $\Delta_{SO}$ and $\mu_B$ denoting the spin-orbit splitting energy and the Bohr magneton, respectively. Pair breaking occurs when the effective Zeeman splitting energy $\sim \frac{H_\parallel^2}{H_{SO}} \mu_B$ overcomes the superconducting gap. It thus yields the following estimate for $H_{c20}^{\parallel} \sim \sqrt{H_{SO} H_P}$, which can greatly exceed $H_P$ if $H_{SO} \gg H_P$ for strong SOIs.

To analyze the entire $H_{c2}^{\parallel} - T_C$ phase diagram, we introduce the pair breaking equation[19] for monolayer NbSe$_2$ with the spin-momentum locking effect incorporated

$$\ln\left(\frac{T_C}{T_{C0}}\right) + \psi\left(\frac{1}{2} + \frac{\mu_B H_\parallel^2/H_{SO}}{2\pi k_B T_C}\right) - \psi\left(\frac{1}{2}\right) = 0. \qquad (1)$$

Here $\psi(x)$ is the digamma function and $k_B$ is the Boltzmann constant. Near $T_{C0}$, Eqn. 1 can be reduced to $\left(1 - \frac{T_C}{T_{C0}}\right) = \frac{(H_{c2}^{\parallel})^2}{H_{SO} H_P}$, which describes well the observed square-root dependence of $H_{c2}^{\parallel} \propto \sqrt{T_{C0} - T_C}$. We fit the experimental $H_{c2}^{\parallel} - T_C$ dependence to the solution of Eqn. 1 with $H_{SO}$ as a free parameter. The best fit (blue line, fig. 4) gives the effective magnetic field $H_{SO} \approx 660$ T, or equivalently, the total spin splitting energy $2\Delta_{SO} \approx 76$ meV. This value agrees well with the values from *ab initio* calculations for the Fermi surface around the K and K' point of the Brillouin zone (~ 70-80 meV)[18], where effects of SOIs on Cooper pairing are the strongest. The upper critical field is determined from Eqn. 1 $H_{c20}^{\parallel} \approx 35$ T, which is over 6 times of $H_P$. To verify this value, we have performed independent differential conductance measurements of a monolayer at 0.3 K up to $H_\parallel$ = 31.5 T. (See Supplementary Materials section 7 for details.) A zero-bias peak originated from Andreev reflections at the normal metal-superconductor contact was observed. With increasing magnetic field, the zero-bias peak diminishes continuously, corresponding to a shrinking superconducting gap, and becomes very weak at 31.5 T (inset of Fig. 4). This result suggests that $H_{c20}^\parallel \gtrsim$ 31.5 T, consistent with the result from Eqn. 1.

Finally, we briefly comment on the $H_{c2}^\parallel - T_C$ phase diagram for the few-layer and bulk sample. As *N* increases, interlayer-coupling kicks in, the orbital effect can quench superconductivity in addition to the spin effect, and the upper critical field $H_{c20}^\parallel$ decreases. However, for layer thickness smaller than the out-of-plane penetration depth (~ 23 nm[10]), the orbital effect is strongly suppressed[19] and similar $H_{c2}^\parallel - T_C$ dependences are observed for few-layer samples with $H_{c20}^\parallel > 3H_P$ in both bilayer and trilayer NbSe$_2$. The observation of $H_{c20}^\parallel \gg H_P$ in bilayer NbSe$_2$, which is centrosymmetric and for which spin-momentum locking is destroyed, suggests the importance of SOIs in bilayers as well. Indeed, as long as interlayer coupling does not cause spin-flip [demonstrated by spin polarized angle resolved photoemission spectroscopy (ARPES) in group-VI TMDs[7]], the out-of-plane spin is still a good quantum number. It is locked to each individual layer, known as spin-layer locking[6], which also protects superconductivity under a parallel magnetic field. On the other hand, in the bulk limit, the orbital effect dominates the phase diagram as for the case of out-of-plane fields. The estimated in-plane upper critical field $H_{c20}^\parallel \approx$ 17 T (from 70% of the linear extrapolation of the $H_{c2}^\parallel - T_C$ dependence at zero temperature[40]), which is about 4 times higher than $H_{c20}^\perp$, is in good agreement with the reported value[9,10]. We note that the above discussion on the magnetic response of few-layer samples is very qualitative. More detailed experimental and theoretical studies are warranted for a full understanding of the interplay between spin-layer locking and orbital effects, and of the importance of the complex Fermi surface of NbSe$_2$[13,14]. Our studies on the strong SOIs and spin-momentum locking in atomically thin NbSe$_2$ down to the monolayer thickness open up a new avenue for the study of interacting electrons with Ising spin in the exact 2D limit.

**Acknowledgments:** We thank Moses H. W. Chan for fruitful discussions.

**Methods:**

**Sample preparation and device fabrication.** High-quality 2H-NbSe$_2$ single crystals were grown from Nb metal wires of 99.95% purity and Se pellets of 99.999% purity by iodine 99.8% vapor transport in a gradient of 730 °C – 700 °C in a sealed quartz tubes for 21 days. A very slight excess of Se was introduced (typically 0.2% of the charge) to ensure stoichiometry in the resulting crystals. Thin flakes were mechanically exfoliated from bulk single crystals on silicone elastomer polydimethylsiloxane (PDMS) stamps. Atomically thin samples of good geometry were first identified by optical microscopy and then transferred onto silicon substrates (covered by a 280 nm layer of thermal oxide) with pre-patterned Au electrodes. To minimize the environmental effects on the samples, we have limited their exposure to air to < 1 hour. Hexagonal boron nitride (h-BN) thin films of 10 – 20 nm thickness were introduced as a capping layer for further protection. The sample thickness was determined according to their shear mode frequency by Raman spectroscopy[17]. The crystal quality was characterized by polarized optical second harmonic generation. (See Supplementary Materials Section 1 for more details).

**Electrical characterization.** Transport and magneto-transport measurements were carried out in a Physical Property Measurement System (PPMS) down to 2.1 K and up to 9 T. For higher magnetic field measurements up to 31.5 T, a Janis He3 cryostat with base temperature of 0.3 K was employed. Unless specifically mentioned, longitudinal electrical resistance was acquired using a four-point geometry with excitation current limited to 1 $\mu A$ to avoid heating and high-bias effects. (Dependence on the excitation current was performed to study the fluctuation effects in 2D.) The devices were mounted on a rotation stick, which allows alignment of the sample plane with the external magnetic field with high accuracy (< 0.5° error). Multiple devices were prepared and measured. All yielded consistent results for samples of the same thickness.

**Figures and Figure Captions:**

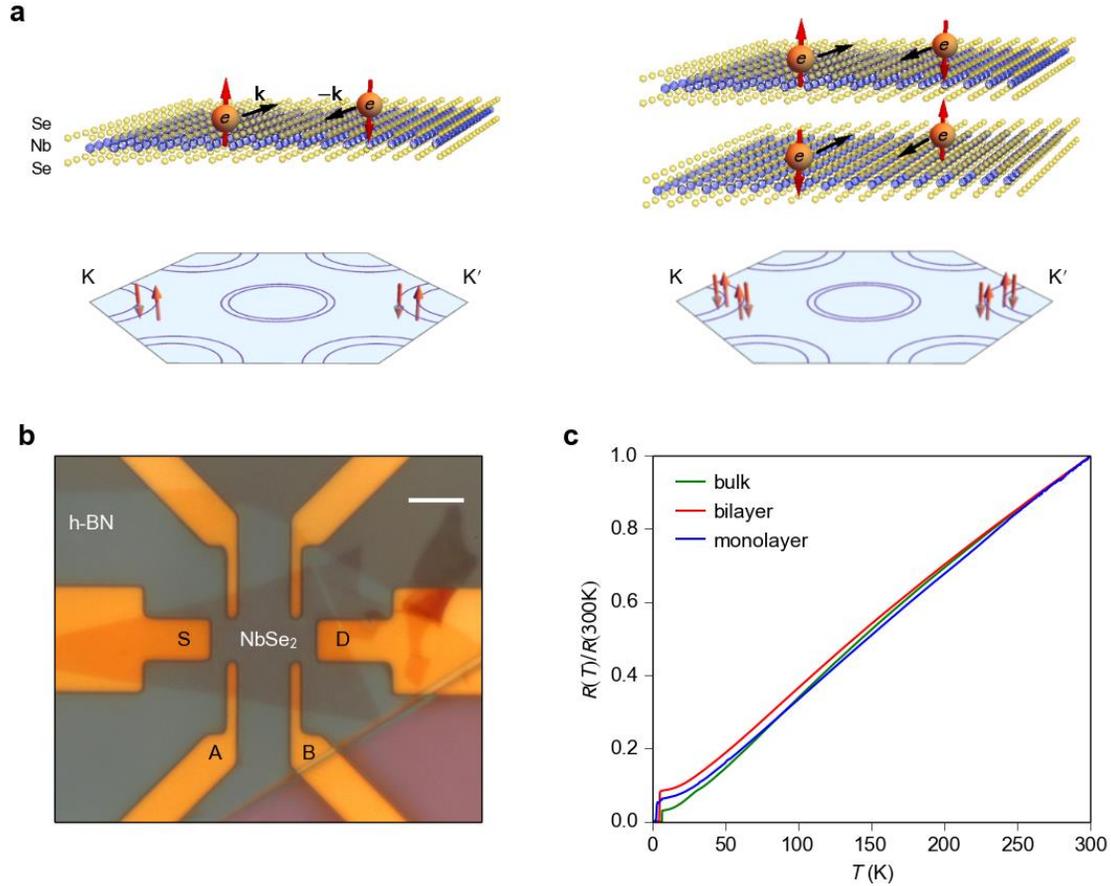

**Fig. 1 Spin-momentum and spin-layer locking in NbSe$_2$.** (**a**) Schematic illustration of spin-momentum locking in monolayer (left) and spin-layer locking in bilayer/bulk (right) NbSe$_2$. First row: Monolayer NbSe$_2$ consists of a layer of Nb atoms (blue balls) sandwiched between two layers of Se atoms (yellow balls) in the trigonal prismatic structure. Bulk 2H-NbSe$_2$ is made of monolayers stacked in the ABAB… sequence. Second row: Brillouin zone (BZ) in the in-plane direction and Fermi surface near the Γ, K and K' point. The Fermi surface is spin split in the monolayer and spin degenerate in the bilayer/bulk. (**b**) Optical image of a bilayer NbSe$_2$ device capped by a thin h-BN layer for environmental protection. The scale bar corresponds to 5 μm. Current was excited through electrode S and D; voltage drop was measured across A and B. (**c**) Temperature dependence of the normalized four-point resistance $R$ for a typical bulk, bilayer and monolayer NbSe$_2$ device from 2.1 – 300 K.

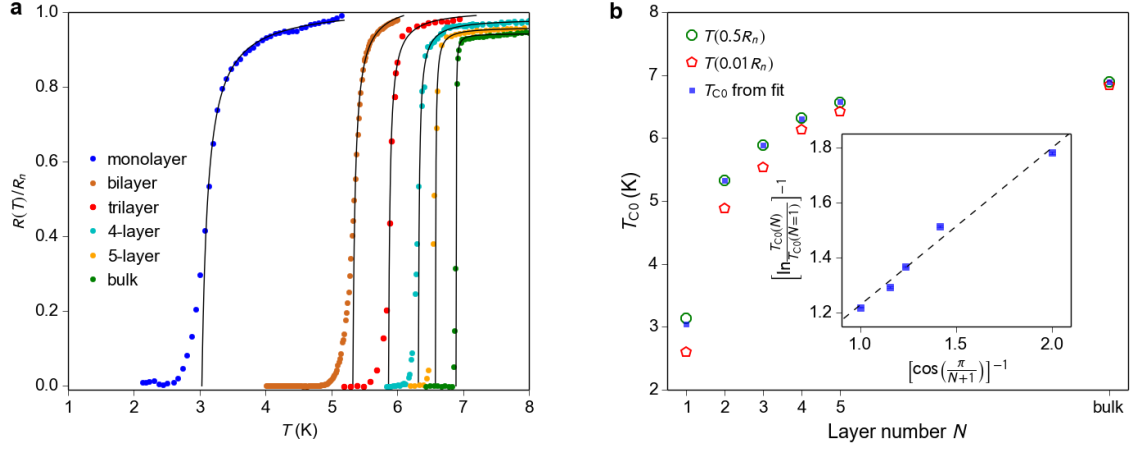

**Fig. 2 Layer number dependence of superconductivity in NbSe$_2$. (a)** Temperature dependence of the resistance for NbSe$_2$ samples of varying thickness. The resistance is normalized to the normal state value right above the superconducting transition. The solid lines are fits to the Aslamasov-Larkin formula for $R(T)/R_n > 0.5$. **(b)** Layer number $N$ dependence of $T(0.5R_n)$, $T(0.01R_n)$, and $T_{C0}$. The inset shows the dependence of $\left[\ln\frac{T_{C0}(N)}{T_{C0}(N=1)}\right]^{-1}$ on $\left[\cos(\frac{\pi}{N+1})\right]^{-1}$ (symbols), where $T_{C0}(N=1) \approx 3.0$ K. Dashed line is the best fit to a linear dependence.

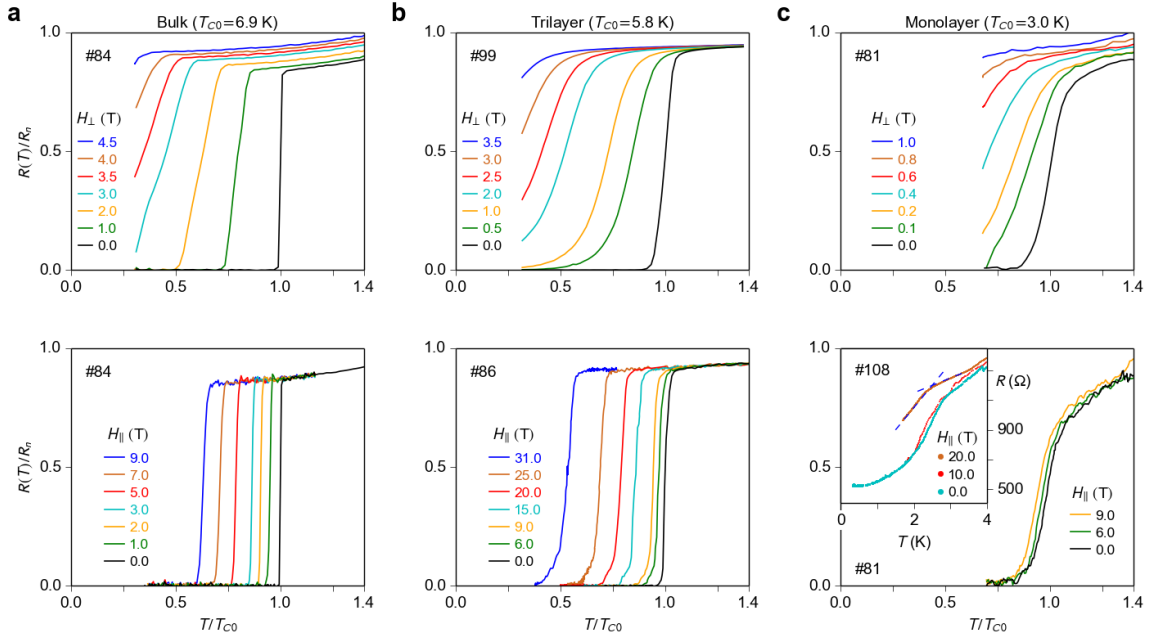

**Fig. 3 Superconductivity of bulk, trilayer, and monolayer NbSe$_2$ under a magnetic field.** Temperature dependence of the dc resistance for a bulk (**a**), trilayer (**b**), and monolayer NbSe$_2$ device (**c**) under out-of-plane ($H_\perp$) and in-plane ($H_\parallel$) magnetic fields. The inset to (**c**) shows the two-point resistance of second monolayer device under in-plane magnetic fields. The crossing of the two dashed lines is used to determine $T_C$.

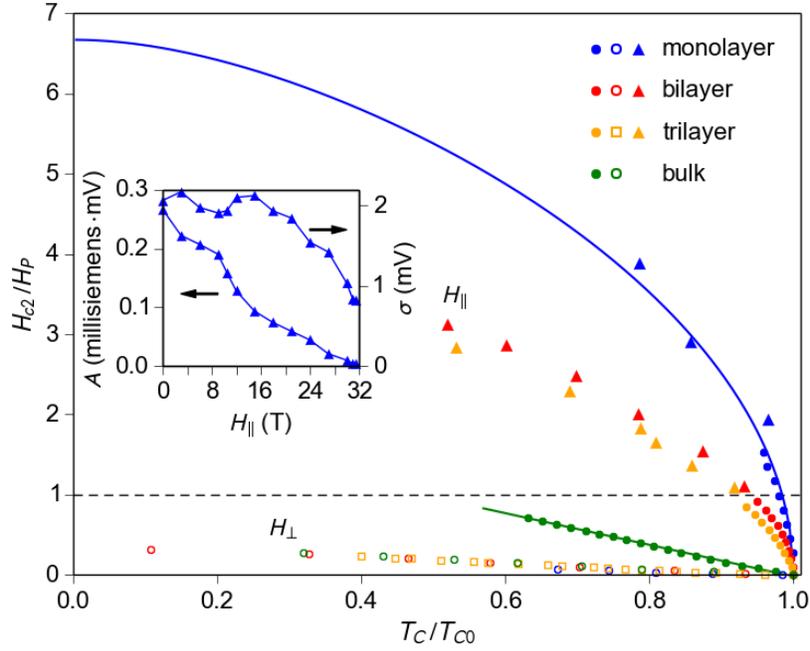

**Fig. 4 $H - T$ superconducting phase diagram for atomically thin NbSe$_2$.** The critical field $H_{c2}/H_P$ as a function of transition temperature $T_C/T_{C0}$ is shown for NbSe$_2$ samples of differing thickness under both out-of-plane $H_\perp$ (open symbols) and in-plane $H_\parallel$ (filled symbols) magnetic fields. The dashed line corresponds to the Pauli paramagnetic limit $H_P$. The blue line is the best fit to the solution of the pair breaking equation. The green line is a linear fit. The inset shows the zero-bias peak area and width as a function of $H_\parallel$ at 0.36 K obtained from differential conductance measurements in a monolayer device. The result shows that superconductivity in monolayer NbSe$_2$ survives under $H_\parallel = 31.5$ T.